\documentclass[11pt,a4wide]{article}
\usepackage{newlfont}
\usepackage{amsmath}
\usepackage{graphics}
\usepackage{float}
\usepackage{graphicx}
\usepackage{epsfig}
\usepackage{multirow}
\usepackage{multicol}
\usepackage{verbatim}
\usepackage{pdflscape}
\usepackage{array}

\begin{document}

\title{ Spectrum of Hybrid Charmonium, Bottomonium and $B_c$ Mesons by Power Series Method}
\author{Nosheen Akbar\thanks{e mail: nosheenakbar@cuilahore.edu.pk}, Zeeshan Ali$^\ddag$, Saadia Arshad$^\ddag$  \\
\textit{$\ast$ Department of Physics, COMSATS University Islamabad, Lahore Campus,}\\
{Lahore, Pakistan.}\\
\textit{$\ddag$ Department of Mathematics, COMSATS University Islamabad, Lahore Campus,}\\
{Lahore, Pakistan.}}
\date{}
\maketitle

\baselineskip=0.30in
\section*{Abstract}

Power series method (PSM) is revisited to find the masses of S, P, and D states of conventional charmonium ($c\overline{c}$), bottomonium ($b\overline{b}$), and $B_c$ ($\overline{b}c$) mesons by assuming the solution of N-dimensional radial Schrodinger equation in series form. An extension in the potential model is proposed by fitting it with lattice data to study the hybrid mesons. The proposed potential model is used to find the masses of hybrid $c\overline{c}$, $b\overline{b}$, and $\overline{b}c$ mesons by applying the power series method. Calculated results are compared with theoretical findings and available experimental data. Our results can be helpful for the investigation of newly experimentally discovered charmonium, bottomonium, and $B_c$ states.

\section*{1. Introduction}
Mesons are hadronic particles made up of a quark and an antiquark pair bounded by a strong gluonic field. When this gluonic field comes in the excited state, mesons are named as "Hybrid Mesons". Many states of conventional and hybrid mesons have been observed in the experiments. The study of these experimentally discovered states of mesons becomes more interesting for physicists. In the investigation of the conventional and hybrid mesons, Schr\"odinger equation (SE) plays the most important role. Schr\"odinger equation is a differential equation which is mostly used to investigate the atomic structure of matter.

To study mesons, SE is solved by different methods like Exact-Analytical Iteration Method \cite{khokha16}, Numerov’s Discretization Method \cite{Allosh}, Fourier grid Hamiltonian method \cite{Allosh}, Nikiforov-Uvarov method \cite{Abushadey16,HMansour2022,Inyang}, Exact Quantization Rule technique \cite{Etido23}, Laplace Transform method \cite{Shady18,Shady18587}, series expansion method \cite{EInyang2022}. In this work, SE is solved by the power series method \cite{MShady2022} to find the masses of conventional and hybrid mesons for different $J^{P C}$ states. This method has the advantage that it can be applied to the equations containing the arbitrary coefficients and boundary conditions.

To study the mesons, different potential models are used. Yukawa potential is adopted in ref. \cite{EInyang2022} to obtain the approximation solution of SE for energy eigenvalues using the series expansion method. Trigonometric Rosen-Morse Potential (TRMP) is employed in ref. \cite{MShady2022} to calculate the mass spectra(MS) of heavy and heavy-light meson in the framework of N-radial fractional Schr\"odinger equation (SE) by using the generalized fractional extended Nikiforov–Uvarov method (GFD-ENU). Cornell plus inverse quadratic potential is used in ref. \cite{Ahmadov} to find the energy spectrum of $c\overline{c}$ and $b\overline{b}$ at finite and zero temperatures. In ref. \cite{MShady2022}, Cornell plus quadratic plus inverse quadratic potential is used to find the mass of bottomonium, charmonium, and $B_c$ mesons by applying the power series method. In present work, we revisit the work of ref. \cite{MShady2022} and then extend this work to find the spectrum of hybrid bottomonium, charmonium and $B_c$ mesons by modifying the potential with the addition of a potential term ($A (1- B r^2))$ whose parameters are determined by fitting this term with the lattice data \cite{Morningstar}. Earlier, this additional term is suggested in the Gaussian form in ref. \cite{Nosheen11}, but extension in gaussian form is not suitable for the power series method.

Hybrid mesons have been studied by different methods like the bag model \cite{bag}, the Lattice QCD \cite{Liu12,Lacock97,Chen21,Dudek13}, the flux tube model \cite{Isgur85,Isgur,Burns}, and QCD sum rules \cite{Balitsky,Chen22,Chen13,Chen,Chen14}. We compare our calculated results with available theoretical and experimental results.

The paper is organized as follows: In section 2, the N dimensional radial Schrodinger equation is described for mesons along with the series solution. Potential models for conventional and  hybrid mesons are defined. Energy expressions derived for conventional as well as hybrid mesons are also described in this section. Section 3 describes the method for finding the values of parameters of the potential models (for conventional and hybrid mesons) for the calculation of masses of charmonium, bottomonium, and $B_c$ mesons. In section 4, the results are discussed in detail.

\section* {2. Solution of Schrodinger Equation and Energy of Mesons by Power Series Method}

N-dimensional radial SE for two interacting particles can be written as
\begin{equation}
\bigg[\frac{d^{2}}{dr^{2}}+\frac{N-1}{r}\frac{d}{dr}-\frac{L(L+N-2)}{r^{2}}+2\mu(E-V(r))\bigg]R(r)=0, \label{E1}
\end{equation}
where $L$ is the angular momentum quantum number, N denotes dimension, and $\mu$ is the reduced mass of the two interacting particles. In this work, these two interacting particles are quark and antiquark, so reduced mass can be defined as $\mu=\frac{m_{q}m_{\overline{q}}}{m_{q}+m_{\overline{q}}}$ with $m_{q}$ and $m_{\overline{q}}$ are the quark and anti-quark masses respectively. $E$ is the energy eigenvalues corresponding to the radial eigenfunctions $R(r)$ and it is approximated in ref. \cite{Kumar,abushadey19} as:
\begin{equation}
R(r)=e^{({-\alpha r^{2} - \beta r})}F(r), \label{E2}
\end{equation}
where $F(r)$ is considered in the series form as
\begin{equation}
F(r)=\Sigma_{n=0}^{\infty}a_{n} r^{\frac{3n}{2}+L}. \label{E3}
\end{equation}
$a_{n}$ is the expansion coefficient. Values of parameters, $\alpha$, $\beta$ depend on the potential model ($V(r)$). Details of the potential V(r) are discussed in next sections 2.1. For the study of hybrid mesons, the potential model used for conventional mesons is extended (discussed below in section 2.2. in detail). Due to the change in the potential model, values of parameters of the radial wave function for hybrids are changed and can be written as:
\begin{equation}
R(r)=e^{({-\alpha' r^{2} - \beta' r})}F(r). \label{E2}
\end{equation}
$\alpha$, $\beta$, $\alpha'$, $\beta'$ are defined below in section 2.1 and 2.2.
\subsection*{2.1. Potential model for conventional meson}

In eq.(\ref{E1}), $V(r)$ represents the potential of interacting particles. For conventional mesons, the potential is modelled\cite{MShady2022,abushadey19} as:
\begin{equation}
V(r)= a r^{2} + b r - \frac{c}{r} + \frac{d}{r^{2}}\label{pot}.
\end{equation}
To find the energy of mesons, we applied the power series method as used in ref.\cite{MShady2022}. From  eq.(\ref{E1}),eq.(\ref{E3})-eq.(\ref{pot}), we obtained the following expressions similar to ref. \cite{abushadey19}.

\begin{equation}
a=\frac{2 \alpha^{2}}{\mu},  \quad \Rightarrow \quad \alpha = \sqrt{\frac{a \mu}{2}}, \label{alpha}
\end{equation}

\begin{equation}
b=\frac{2 \alpha \beta}{\mu},  \quad \Rightarrow \quad \beta = \frac{b \mu}{2 \alpha}, \label{beta}
\end{equation}

\begin{equation}
c=\frac{-\beta(3n+2L+2)-(N-1) \beta}{2\mu}, \quad \Rightarrow \quad \beta = \frac{- 2 \mu c}{3n + 2L + N +1}, \label{c}
\end{equation}
\begin{equation}
d=\frac{L(L+N+2)-2(N-1)(3n+2L+2)-4L(L+N+2)}{8 \mu}
\end{equation}
and
\begin{equation}
E =\frac{\alpha (3n+2L+N)}{\mu}- \frac{\beta^{2}}{2 \mu} \label{E}
\end{equation}
Eq.(\ref{E}) can also be written as
\begin{equation}
E =\frac{a}{2 \mu} (3n+2L+N)- \frac{b^2}{4 a} \label{EE}
\end{equation}

\subsection*{2.2. Potential model for hybrid meson}

To study the hybrid mesons, above mentioned potential model (defined in eq.\ref{pot}) is modified as:
\begin{equation}
V(r)= ar^{2}+br-\frac{c}{r}+\frac{d}{r^{2}} + A (1 - B r^2). \label{hp}
\end{equation}

Here, the additional term $(A (1 - B r^2))$ is suggested for the potential energy difference between conventional and hybrid mesons. The parameters $A$ and $B$ are found by fitting $(A (1 - B r^2))$ with the lattice simulation data obtained from Fig. 3 of ref.\cite{Morningstar} for the potential energy differences ($\upsilon_{d}$) between the ground and excited states. The best fit is obtained for $A = 1.40498 GeV$ and $B=0.016824 GeV^3 $. $\chi^2$ is calculated by using the following formula:
\begin{equation}
\chi^{2} = \frac{\sum^{^{n}}_{_{i=1}}[\upsilon_{d}(r_i) - A (1 - B r_i^2)]}{\sum^{^{n}}_{_{i=1}}(\upsilon_{d}(r_i))^2},\label{P26}
\end{equation}
In the flux tube model \cite{Isgur85}, $\frac{\pi}{r}$ is used for the excited part of the quark-antiquark potential. Few models in the Gaussian form for the potential energy differences between ground and excited states are suggested in ref. \cite{Nosheen11}. A comparison of $\chi^2$ for these models with our newly used potential model is reported in Table \ref{chi2}. It is observed that our newly suggested potential model has $\chi^2$ less than $\pi/r$.
\begin{table}
\caption{\label{chi2} $\chi^{2}$ for the lattice data(for potential difference) with models ($V_g (r)$) with best fit parameter's values.}
\begin{center}
\begin{tabular}{|c|c|c|c|c|}
\hline
& &\multicolumn{3}{|c|}{Parameters} \\
$ans\ddot{a}tz$ &$\chi^{2}$ & $A$ & $B$ & $c$ \\ \hline
& & GeV &  &  \\ \hline
$A (1- B r^2)$ & 0.1459 & 1.40498 &0.016824 $GeV^2$& - \\
$\pi/r$ & 0.2305 & - & - & - \\
$A \times \textrm{exp} (- B r^2)$ & 0.0857 & 1.8139 & 0.0657 $GeV^2$ & -  \\
$\frac{c}{r} + A \times \textrm{exp} (- B r)$ & 0.0012 & 1.2448 & 0.1771 GeV& 0.3583 \\  \hline
\end{tabular}
\end{center}
\end{table}

With the modified potential model for hybrid mesons defined in eq.(\ref{hp}) , SE can be written as:
 \begin{multline}
\Sigma_{n=0}^{\infty}a_{n}\bigg[\bigg(\beta'^{2}+4\alpha'^{2}r^{2}+4\alpha'\beta' r-2\alpha' +\frac{N-1}{r}(-2\alpha' r - \beta') -\frac{L(L+N-2)}{r^{2}}+2\mu E-2\mu A (1- B r^{2})\bigg)r^{\frac{3 n}{2}+L}\\+\bigg((-2\beta-4\alpha r)(\frac{3n+2L}{2})+\frac{N-1}{r}(\frac{3n + 2L}{2})+2\mu c \bigg) r^{\frac{3 n}{2}+L-1}+\bigg(\frac{(3n+2L-2)(3n+2L)}{4}-2\mu d\bigg)r^{\frac{3 n}{2}+L-2}\\-2\mu a r^{\frac{3 n}{2}+L+2}-2\mu b r^{\frac{3 n}{2}+L+1}\bigg]=0
\end{multline}
equating each coefficient of r to zero, we obtained the following relations:
\begin{equation}
E = A + \frac{\alpha'}{\mu}(3n + 2L + N)- \frac{\beta'^{2}}{2 \mu}
\end{equation}
\begin{equation}
\alpha'=\sqrt\frac{\mu (a- A B)}{2}, \label{alphah}
\end{equation}
\begin{equation}
\beta'=\frac{\mu b}{2 \alpha'}, \quad \quad \beta' = \frac{- 2 \mu c}{3n + 2L + N +1} \label{betah}
\end{equation}

\section*{3. Parameters of Potential Model and Mass spectra of mesons}
\subsection*{3.1. Conventional mesons}

Mass of the mesons can be calculated by adding the constituent quark masses in the energy(E) defined in eq.(\ref{EE}), i.e;
\begin{equation}
 M = m_{q} +m_{\overline{q}} + \frac{a}{2 \mu} (3n+2L+N)- \frac{b^2}{4 a}. \label{fitt}
\end{equation}
This equation helps in finding the parameters ($\alpha, \beta, a,b,c,d$) by taking M equal to the experimental mass for a particular meson ($c\overline{c}$, $b\overline{b}$ and $b\overline{c}$).

To find the potential parameters a and b, charm and bottom quark mass is taken equal to 1.48 $GeV$ and 4.75 $GeV$ respectively. In case of charmonium, two algebraic equations are obtained by substituting the experimental values of $1S$ and $2D$ mass in eq.(\ref{fitt}). These two equations are solved for finding the value of parameter $a$ and $b$. Substitution of these values of $a$ and $b$ in eqs.(\ref{alpha},\ref{beta}), $\alpha$ and $\beta$ are calculated. With this calculated value of $\beta$, eq.(\ref{c}) gives the value of parameter $c$.

For bottomonium, same steps are repeated to find $a$, $b$, $\alpha$, $\beta$ and $c$ by inserting experimental values of M for $1 S$ and $3 P$ in eq.(\ref{fitt}). For $B_c$, two algebraic equations are obtained by inserting experimental values of M for $1 S$ and $2 S$ in eq.(\ref{fitt}).

Substituting the values of parameters for each sector in eq.(\ref{fitt}), we find the mass of different states of mesons by varying quantum numbers ($n,L$) and taking N=3.

\subsection*{3.2. Hybrid mesons}

Mass of hybrid mesons is calculated by the following relation:
\begin{equation}
M =  m_{q} +m_{\overline{q}}+ A + \frac{\alpha'}{\mu}(3n + 2L + N)- \frac{\beta'^{2}}{2 \mu}. \label{mass}
\end{equation}
The parameters $\alpha'$ and $\beta'$ are obtained from eqs.(\ref{alphah},\ref{betah}) where $a,b$, and $c$ are the parameters of the conventional meson potential model. By taking N=3, and using the calculated values of all parameters, masses are calculated for different excited states of hybrid meson by varying $n,L$.

\section*{4. Results and Discussion}

In the present work, we derive the expressions for the energy of conventional and hybrid mesons by solving the N-dimensional radical Schr\"odinger equation using the power series technique. By using these energy expressions, masses of conventional and hybrid charmonium, bottomonium, and $B_c$ mesons are calculated for ground, radial, and orbital excited states. Our calculated masses of conventional $c\overline{c}$, $b\overline{b}$, and $\overline{b}c$ mesons are reported in Tables 2-4 along with the experimental and other theoretical calculated masses.

\begin{table}[h!]
\begin{center}
\caption{\label{Table1}Mass spectra of $c\overline{c}$ in GeV with a=0.0341$\textrm{GeV}^{3}$ and b = 0.20826 $\textrm{GeV}^{2}$ }
\begin{tabular}{ |c|c|c|c|c|c| }
 \hline
 States & This work & \cite{abushadey19} &  \cite{Nosheen11} & \cite{Barnes05}&   Exp \cite{pdg23}\\ \hline
 1S     & 3.0974 & 3.068    & 3.09   &     3.09&   3.0969 $\pm$ 0.000006 \\
 2S     & 3.5528 & 3.663     & 3.6718&    3.672&    3.6861 $\pm$ 0.000025   \\
 3S     & 4.0081 & 4.258    &   4.0716&   4.072& -\\
 4S     & 4.4635 & 4.852    &  4.406  &   4.406& - \\
 5S     & 4.9189 &  -     &    4.7038 &    -   & - \\
 6S     & 4.9189 &  -     &    4.9769 &    -   & -  \\
 1P     & 3.401  &  3.464   &   3.4245&   3.424  &     3.41475 $\pm$ 0.00031 \\
 2P     & 3.8564 & 4.059    &   3.8523 &  3.852& -\\
 3P     & 4.3117 &    -    &    4.2017&   4.202&  - \\
 4P     & 4.7671&     -     &   4.5092&   -    &   -\\
 5P     & 5.2225 &    -    &    4.7894&   -    & - \\
 1D     & 3.7046 &  3.861      & 3.7850&  3.785 &        3.77313 $\pm$ 0.00035 \\
 2D     & 4.1599 &    -     &    4.1415 & 4.142 &         4.159 $\pm$ 0.00021 \\
 3D     & 4.6153&     -     &     4.4547 & -    & -\\
 4D     & 5.0707&     -      &     4.7395&  -   &  -\\
 \hline
\end{tabular}
\end{center}
\end{table}

\begin{table}[h!]
\begin{center}
\caption{\label{Table2} Mass spectra of $b\overline{b}$ in GeV with  a=0.0936$GeV^{3}$ and b= 0.4151$GeV^{2}$}
\begin{tabular}{ |c|c|c|c|c|c|c|}
 \hline
 States & This work  &  \cite{abushadey19}  & \cite{Nosheen17} & \cite{14110585}&  Exp \cite{pdg23}  \\ \hline
 1S     & 9.4609 & 9.46  &     9.5299   & 9.459&  9.4603 $\pm$ 0.00026  \\
 2S     & 9.8820  & 10.023&   10.010  &   10.004& 10.023 $\pm$ 0.00031 \\
 3S     & 10.3032 & 10.585&   10.295  &   10.354& 10.3552 $\pm$ 0.0005   \\
 4S     & 10.7243 & 11.148&   10.5244 &    10.663 & 10.5794 $\pm$ 0.0012  \\
 5S     & 11.1454 & -    &     10.7251 &   10.875 & 10.8852 $\pm$ 0.00016  \\
 6S     & 11.5665& -     &    10.9074  &     -    &   -                   \\
 1P     &  9.7416& 9.8354     &9.9326 &    9.896 &   9.91221 $\pm$ 0.00026  \\
 2P     & 10.1628 & 10.398 &   10.2245&    10.261&    10.26865 $\pm$ 0.00022 \\
 3P     & 10.5839&   -     &     10.4585&   10.549&    10.524 $\pm$ 0.0008 \\
 4P     & 11.005& -      &     10.6627&      10.797&   -   \\
 5P     & 11.4262& -       &10.8478&         -&        -   \\
 1D     & 10.0224& 10.210      &     10.1389&10.155&    10.1637 $\pm$0.0014  \\
 2D     & 10.4435& -         &    10.3799& 10.455&      - \\
 3D     & 10.8647& -          &   10.5892& 10.711&      - \\
 4D     & 11.2858& -           &   10.7782& 10.939&     - \\ \hline
\end{tabular}
\end{center}
\end{table}

\begin{table}[h!]
\begin{center}
\caption{\label{Table3} Mass spectra of $B_c$ mesons in GeV with a=0.0806$GeV^{3}$ and b= 0.4104$GeV^{2}$ }
\begin{tabular}{ |c|c|c|c|c|c|c|c|}
 \hline
 States & This work & \cite{abushadey19} & \cite{Nosheen19}&  \cite{godfrey04} & Exp \cite{pdg23} & Lattice \cite{Davies} \\ \hline
 1S     & 6.2745 & 6.277&    6.2749&    6.332&      6.2749$\pm$ 0.008 &  6.280$\pm$0.030$\pm$0.190\\
 2S     & 6.8415 & 6.4963&   6.841&      6.881&      6.842 $\pm$0.004&  6.960$\pm$0.080 \\
 3S     & 7.4084 & 6.7148&    7.197&      7.235&          - &\\
 4S     & 7.9754 & 6.9333&    7.488&      -&             - & \\
 5S     & 8.5423 &   -   &     -&          -&             - &\\
 6S     & 9.1092&    -   &      -&          -&             - &\\
 1P     & 6.6525 & 6.4234&     6.753&        6.734 &       -  &  6.783$\pm$0.030          \\
 2P     & 7.2194 &  6.6419&    7.111&         7.126&       - &-\\
 3P     & 7.7864 &   -    &    7.406&         -    &       -  & -     \\
 4P     & 8.3533 &    -    &   -    &          -&          -  &-\\
 5P     & 9.4872&     -   &    -    &          -&           -   &-              \\
 1D     & 7.0304  & 6.569 &   6.998&           7.072&       -   &-      \\
 2D     & 7.5974 &  -     &   7.302&              -&        -     &-       \\
 3D     & 8.1643&  -      &   7.57&               -&        -       &-         \\
 4D     & 8.7313 & -      &   -&                 -&         -         &-    \\

 \hline
\end{tabular}
\end{center}
\end{table}

Masses of hybrid $c\overline{c}$, $b\overline{b}$, and $\overline{b}c$ mesons are reported in Tables 5,6 along with the masses calculated by others by different methods. By observing the results reported in Tables 2-6, it is concluded that mass is increasing toward higher states. It is also observed that hybrid mesons are heavier than the normal mesons for the same quantum numbers (n,L). Masses of hybrid charmonium for lowest and first excited $J^{P C}$ states are compared with others theoretical work in Tables 8,9. Comparison of hybrid bottomonium and $B_c$ for the lowest $J^{P C}$ states is given in Tables 10,11. Here, $J =L\oplus S$, $P$ is the parity and $C$ is the charge quantum numbers defined as $P = (-1)^{L+1}$ and $C = (-1)^{L+S}$. $L$  and $S$ are the total angular momentum and the spin angular momentum quantum numbers for meson. For hybrid mesons $P$ and $C$ are defined as $P = \epsilon (-1)^{L+\Lambda +1}$, and $C = \epsilon \eta
(-1)^{L + \Lambda + S}$ with $\epsilon, \eta = \pm 1$. where $\Lambda = 0$ for conventional mesons and $\Lambda = 1$ for hybrid mesons \cite{Kuti97}. In the present work, spin interactions are not incorporated in the potential model so the charge quantum number (C) remains unchanged for the same value of $L$ with $S=0,1$. Following are a few observations obtained from the tables for hybrid mesons:
\subsection*{Hybrid Charmonium}
Hybrid charmonium states with $J^{P C}= 0^{++}, 1^{+-}, 0^{--}, 1^{-+}$ have the same mass equal to 4.2992 GeV which is the lowest calculated mass for hybrid charmonium. This value of lowest mass is comparable to the numerically reported lowest mass in ref. \cite{Nosheen11} with $J^{P C}= 0^{--}, 0^{++}$. In Lattice QCD \cite{Liu12}, the lowest mass is 4.189 GeV with $J^{P C}= 1^{--}$. Recently, ref.\cite{2024} reported the lowest mass value equal to 3.56 with $J^{P C}= 0^{-+}$. From these observations, we conclude that our prediction for the lowest mass is much closer to the predictions of refs.\cite{Liu12,Nosheen2014}.

\subsection*{Hybrid Bottomonium}
For hybrid bottomonium, the states with $J^{P C}= 0^{++}, 1^{+-}, 0^{--}, 1^{-+}$ have mass equal to 10.8088 GeV which is the lowest calculated mass. This value of lowest mass is comparable to the numerically reported lowest mass in ref. \cite{Nosheen17} with $J^{P C}= 0^{--}, 0^{++}$. In refs. \cite{Chen13,Chen14u}, the lowest mass is 9.68 GeV with $J^{P C}= 0^{-+}$ while in ref.\cite{iddir}, the lowest mass is 10.5 GeV with  $J^{P C}= 1^{++}$. Ref.\cite{2024} also reported the lowest mass value equal to 9.68 with  $J^{P C}= 0^{+-}$.

\subsection*{Hybrid $B_c$}

In case of hybrid $B_c$ mesons, the lowest mass is calculated as equal to 7.3724 GeV with $J ^P = 1^-$. Refs. \cite{Nosheen19,Chen14} also found the $1^-$ state with the lowest mass equal to 7.422 and 6.83 respectively. Ref.\cite{2024} reported the lowest state mass as 6.63 GeV, which is much smaller than our calculated mass.

\begin{table}[h!]
\begin{center}
\caption{\label{Table 5} Mass spectra of hybrid Charmonium  and hybrid botmonium in (GeV). Masses are rounded to 4 decimal places}
\begin{tabular}{ |c|c|c|c|c|c|c|c|}
\hline
\multicolumn{4}{|c|}{hybrid charmonium}&\multicolumn{4}{|c|}{hybrid botmonium}\\  \hline
 States & This work & NR\cite{Nosheen2014} & Rel. \cite{Nosheen2014} & States &  This work & NR \cite{Nosheen17}& Rel. \cite{Nosheen17}   \\ \hline
1S     & 4.2992 &  4.1063      &    4.1707  & 1S   &  10.8088&    10.7747&     10.8079\\
2S     & 4.5515 &  4.4084     &     4.4837  & 2S   &  11.1729&    10.9211&     10.928\\
3S     & 4.8037 &  4.6855    &      4.7614  & 3S   &  11.537&     11.0664&     11.048\\
4S     & 5.056 &   4.9438  &        5.0132  & 4S   &  11.9011&    11.2086&     11.1662\\
5S     & 5.3082 &  5.1876   &       5.2448  & 5S   &  12.2652&    11.3469&     11.2817 \\
6S     & 5.5604 &  5.4197   &       5.4602&   6S   &  12.6293&    11.4814&     11.394 \\
1P     & 4.4674 &  4.2464    &      4.3203 &  1P   &  11.0516&    10.8366&      10.8569 \\
2P     & 4.7196&   4.5264    &      4.6070&   2P   &   11.4157&   10.9857&      10.9856 \\
3P     & 4.9719 &  4.7875     &     4.8659  & 3P   &   11.7797&   11.1372&      11.1097 \\
4P     & 5.2214 &  5.0338     &     5.1034  & 4P   &   12.1438&   11.2795&      11.2293 \\
5P     & 5.4764 &  5.2682    &      5.3236&   5P   &   12.5079 &   11.5503&      11.4566 \\
1D     & 4.6356 &  4.4232   &       4.4320&   1D   &   11.2943&   10.9063&      10,9127 \\
2D     & 4.8878 &  4.6955    &      4.6892&   2D   &   11.6584&   11.0591&      11.0415 \\
3D     & 5.14 &    4.9402 &         4.966&    3D   &   12.0225&   11.2054&      11.165 \\
4D     & 5.3923  & 5.1912    &      5.2034 &  4D   &   12.3866&   11.3463&      11.2837 \\
 \hline
\end{tabular}
\end{center}
\end{table}

\begin{table}[h!]
\begin{center}
\caption{Mass spectra of hybrid Bc mesons in (GeV). Our calculated masses are rounded to 4 decimal places}
\begin{tabular}{ |c|c|c|c|c|c|c| }
 \hline
 States & This work & \cite{Nosheen19}  & \cite{Chen14}   \\ \hline
$B_{c}^{h}$ 1S     & 7.3724 & 7.415      &  6.90$\pm$0.12$\pm$0.09 or 7.37$\pm$0.12$\pm$0.07$\pm$0.12 \\
$B_{c}^{h}$ 2S     & 7.849 & 7.646      &                     -\\
 $B_{c}^{h}$3S     & 8.3256 & 7.866     &                  -\\
$B_{c}^{h}$ 4S     & 8.8022 & 8.075     &          - \\
 $B_{c}^{h}$5S     & 9.2788 & -     &                -  \\
 $B_{c}^{h}$6S     & 9.7555 & -    &               -  \\
 $B_{c}^{h}$1P     & 7.6901 & 7.547      &   7.15$\pm$0.08$\pm$0.05$\pm$0.09 or 7.67$\pm$0.07$\pm$0.02$\pm$0.12 \\
$B_{c}^{h}$ 2P     & 8.1668 & 7.776     &               - \\
$B_{c}^{h}$ 3P     & 8.6434 & 7.990    & -  \\
$B_{c}^{h}$ 4P     & 9.12&    -   &                  -\\
$B_{c}^{h}$ 5P     & 10.073&  -     &          - \\
$B_{c}^{h}$ 1D     & 8.0079 & 7.663    &        - \\
$B_{c}^{h}$ 2D     & 8.4845 & 7.886     &       -  \\
$B_{c}^{h}$ 3D     & 8.9611 & 8.095    &        -  \\
$B_{c}^{h}$ 4D     & 9.4377 & -   &            -   \\  \hline
\end{tabular}
\end{center}
\end{table}

\begin{table}[h!]
\begin{center}
\caption{\label{Table 7a}  $J^{P C}$ states }
\begin{tabular}{ |c|c|c|c|c|c|c|c| }
 \hline
L & S & $J^{P C}$ for conventional & $J^{P C}$ for Hybrid \\ \hline
0 & 0, 1 & $0^{-+}$, $1^{--}$ &  $0^{++}$, $1^{+-}$  \\
 & & & $0^{--}$, $1^{-+}$ \\ \hline
1 & 0, 1 & $0^{++}$, $1^{+-}$, $1^{++}$ , $2^{++}$ &  $0^{-+}$, $1^{--}$, $1^{-+}$ , $2^{-+}$  \\
 & & &  $0^{+-}$, $1^{++}$, $1^{+-}$ , $2^{+-}$ \\ \hline
2 & 0, 1 & $2^{-+}$, $1^{--}$, $2^{--}$ , $3^{--}$ & $2^{++}$, $1^{+-}$, $2^{+-}$ , $3^{+-}$ \\
 & & & $2^{--}$, $1^{-+}$, $2^{-+}$ , $3^{-+}$ \\ \hline
\end{tabular}
\end{center}
\end{table}

\begin{table}[h!]
\begin{center}
\caption{\label{Table 6a} Lowest Mass $J^{P C}$ states of Hybrid Charmonium}
\begin{tabular}{ |c|c|c|c|c|c|c|c| }
 \hline
 States & This work & \cite{Nosheen2014} & \cite{Isgur85} & Lattice QCD \cite{Liu12}& \cite{2024} \\  \hline
 $1^{+-}$ & 4.2992&     4.1063 &    4.19&         -  &     4.21$\pm$0.15    \\
 $1^{-+}$ & 4.2992 &    4.1063&     4.19&         4.213 &  3.93$\pm$0.10 \\
 $0^{++}$ & 4.2992 &    4.0802&      - &           -    &  4.53$\pm$0.06\\
 $0^{--}$&  4.2992&     4.0802&      - &           -    &   4.63$\pm$0.14\\
 $1^{--}$&  4.4674&     4.2678&       4.19&        4.189 &  4.12$\pm$0.11\\
 $1^{++}$&  4.4674&     4.2678&       4.19&          -    & 4.15$\pm$0.09 \\
 $0^{-+}$&  4.4674&     4.2464&       4.19&         4.920  & 3.56$\pm$0.09 \\
 $0^{+-}$&  4.4674&     4.2464&       4.19 &        4.35  &  4.06$\pm$0.12 \\
 $2^{-+}$&  4.4674&     4.2739 &      4.19  &       4.3 & -     \\
 $2^{+-}$&  4.4674&     4.2739&       4.19&         4.4 & -   \\
 $3^{+-}$&  4.6356&     4.4197&       -&             -   & -  \\
 $3^{-+}$&  4.6356&     4.4197&        -&            4.7  & - \\
 \hline
\end{tabular}
\end{center}
\end{table}

\begin{table}[h!]
\begin{center}
\caption{\label{Table 6b}  Mass of first excited $J^{P C}$ states of Hybrid Charmonium}
\begin{tabular}{ |c|c|c|c| }  \hline
 States & This work & \cite{Nosheen2014} & Lattice QCD \cite{Liu12}\\  \hline
 $1^{+-}$ & 4.5515&     4.4084 & - \\
 $1^{-+}$ & 4.5515 &    4.4080& - \\
 $0^{-+}$ & 4.7196 &    4.5264& - \\
 $0^{+-}$&  4.7196&     4.5264& -  \\
 $2^{-+}$&  4.7196&     4.5264 & -\\
 $2^{+-}$&  4.7196&     4.5653& 4.505  \\
  \hline
\end{tabular}
\end{center}
\end{table}

\begin{table}[h!]
\begin{center}
\caption{\label{Table7}  Masses of Hybrid Bottomonium for different $J^{P C}$ states with lowest mass }
\begin{tabular}{ |c|c|c|c|c|c|c|}
 \hline
 States & This work & \cite{Nosheen17}  &  QCD sum rules \cite{Chen14u}\cite{Chen13} & \cite{iddir} & \cite{2024} \\ \hline
 $0^{--}$ & 10.8088&     10.8069 &         11.48$\pm$0.75 &   10.66 &          10.51$\pm$0.03  \\
 $0^{++}$ & 10.8088&     10.8069&          11.20$\pm$0.48&    -     &          10.57$\pm$0.08 \\
 $1^{+-}$ & 10.8088&     10.8079&          10.70$\pm$0.53&    -     &          10.46$\pm$0.06\\
 $1^{-+}$&  10.8088&     10.8079&          9.79$\pm$0.22&    10.80  &          9.85$\pm$0.11\\
 $1^{--}$&  11.0516&     10.8561&          9.7$\pm$0.12&       -   & - \\
 $1^{++}$&  11.0516&     10.8561&          11.09$\pm$0.60&    10.50&           10.41$\pm$0.18     \\
 $0^{-+}$&  11.0516&     10.8534&          9.68$\pm$0.29&      -   &           10.55$\pm$0.10   \\
 $0^{+-}$&  11.0516&     10.8534&          10.17$\pm$0.22&    10.68&           9.68$\pm$0.20    \\
 $2^{-+}$&  11.0516&     10.8569&          9.93$\pm$0.21 &     -  &            10.12$\pm$0.06 \\
 $2^{+-}$&  11.0516&     10.8569&             -&               -& \\
 $3^{+-}$&  11.2943&     10.9127&             -&              -& - \\
 $3^{-+}$&  11.2943&     10.9127&             -&              -& -  \\
 $2^{++}$&  11.2943&     10.9125&          10.64$\pm$0.03  &   & -  \\
 $2^{--}$&  11.2943&     10.9125&             -&                & - \\
 \hline
\end{tabular}
\end{center}
\end{table}

\begin{table}[h!]
\caption{\label{Table9}  Masses of Hybrid $B_c$ for different $J^{P C}$ states with lowest mass}
\begin{center}
\begin{tabular}{ |c|c|c|c|c|c|c|}
 \hline
 States &    P.W &     \cite{Nosheen19} & QCD \cite{Chen14} &\cite{2024} \\ \hline
 $1^{-}$ & 7.3724&     7.422 &     6.83$\pm$0.08$\pm$0.01$\pm$0.07&        6.63$\pm$0.14 \\
 $1^{+}$ & 7.3724 &    7.422&      7.7$\pm$0.06$\pm$0.05$\pm$0.13&          7.17$\pm$0.09 \\
 $0^{-}$ & 7.3724 &    7.415&       6.90$\pm$0.12$\pm$0.01$\pm$0.09&         6.65$\pm$0.08\\
 $0^{-}$&  7.3724&     7.415&       7.73$\pm$0.12$\pm$0.07$\pm$0.12 &        7.03$\pm$0.12\\
 $2^{-}$&  7.6901&     7.547 &       7.15$\pm$0.08$\pm$0.05$\pm$0.09  &  - \\
 $2^{+}$&  7.6901&     7.547&        7.67$\pm$0.07$\pm$0.02$\pm$0.09& -\\ \hline
\end{tabular}
\end{center}
\end{table}


\begin{thebibliography}{00}
\bibitem{khokha16}
E. M. Khokha, M. Abu-Shady and T. A. Abdel-Karim, Quarkonium Masses in the N-dimensional Space Using the Analytical Exact Iteration Method, International J. of Theor. and Appl. Maths. Vol. 2, 2016, 86.

\bibitem{Allosh}
M. Allosh, M. S. Abdelaal, F. Alshowaikh and A. Ismail, Numerical Solutions of a Three-Dimensional Schr\"odinger Equation for a Non–Relativistic Quark Model, Appl. Maths. Inf. Sci., Vol. 17, 2023, 447.

\bibitem{Abushadey16}
M. Abu-Shady, Heavy Quarkonia and Bc Mesons in the Cornell Potential with Harmonic Oscillator Potential in the N-dimensional Schr\"odinger Equation, International J. of Appl. Maths. and Theor. Phys., Vol. 2, 2016, 16.

\bibitem{HMansour2022}
H. Mansour and A. Gamal, Meson Spectra using Nikiforov-Uvarov Method, Results in Phys., Vol. 33, 2022, 105203.

\bibitem{Inyang}
E. P. Inyang, A. N. Ikot, E. P. Inyang, I. O. Akpan, J. E. Ntibi, E. Omugbe and E. S. William, Analytic study of thermal properties and masses of heavy mesons with Quarkonium Potential, Results in Phys., Vol. 39, 2022, 105754.

\bibitem{Etido23}
E. P. Inyang, F. O. Faithpraise, J. Amajama, E. S. William,  E. O. Obisung and J. E. Ntibi, Theoratical Investigation Of Meson Spectrum Using Exact Quantization Rule Technique, East Eur. J. Phys., Vol. 1, 2023, 53.

\bibitem{Shady18}
Abu-Shady and M. Khokha, Heavy-Light mesons in the non-relativistic Quark model using Laplace Transformation method with the Generalized Cornell potential, Adv. High Energy Phys., Vol. 2018, 7032041.

\bibitem{Shady18587}
Abu-Shady, M. Abdel-Karim and T. A. Khokha, Exact solution of the N-dimensional radial Schr\"odinger equation via Laplace transformation method with the Generalized Cornell potential. J. Quantum Phys., 2018, 45.

 \bibitem{EInyang2022}
 E. P. Inyang, I. O. Akpan, J. E. Ntibi and E. S. William, Analytical Solutions of the Schr\"odinger Equation with Class of Yukawa Potential for a Quarkonium System Via Series Expansion Method, East Eur. J. of Phys., Vol. 2, 2022, 43.

\bibitem{MShady2022}
  M. Abu-shady and H. M. Fath-Allah, The Effect of Extended Cornell Potential on Heavy and Heavy-Light Meson Masses Using Series Method, East Eur. J. of Phys., Vol. 4, 2022, 80.

\bibitem{Ahmadov}
A. I. Ahmadov, C. Aydin and O. Uzun, Bound state solution of the Schrodinger equation at finite temperature, J. of Phys. Conf. Series, Vol. 1194, 2019, 012001.

\bibitem{Morningstar}
K. J. Juge, J. Kuti and C. Morningstar, The Heavy quark hybrid meson spectrum in lattice QCD, AIP Conf. Proc., Vol. 688, 2003, 193-.

\bibitem{Nosheen11}
M. Atif Sultan, Nosheen Akbar, Bilal Masud, and Faisal Akram, Higher hybrid charmonia in an extended potential model, Phys. Rev., Vol. 90, 2014, 054001.

\bibitem{bag}
 P. Hasenfratz, R. R. Horgan, J. Kuti and J. M. Richard, The effects of coloured glue in the QCD motivated bag of heavy quark-antiquark systems, Phys. Lett. B, Vol. 95, 1980, 299.

\bibitem{Liu12}
 L. Liu, G. Moira, M. Peardona, S. M. Ryana, C. E. Thomasa, P. Vilasecaa, J. J. Dudekb, R. G. Edwards, B. J. David and G. Richards, Excited and exotic charmonium spectroscopy from lattice QCD, J. High Energy Phys., Vol. 2012, 2012, 126.

\bibitem{Lacock97}
 P. Lacock, C. Michael, P. Boyle and P. Rowland, Hybrid mesons from quenched QCD, Phys. Lett. B, Vol. 401, 1997, 308.

\bibitem{Chen21}
 Y. Ma, Y. Chen, M. Gong and Z. Liu, Strangeonium-like hybrids on the lattice, Chin. Phys. C, Vol. 45, 2021, 013112.

\bibitem{Dudek13}
 J. J. Dudek, R. G. Edwards, P. Guo and C. E. Thomas, Toward the excited isoscalar meson spectrum from lattice QCD, Phys. Rev. D, Vol. 88, 2013, 094505.

\bibitem{Isgur85}
 N. Isgur and J. E. Paton, Flux-tube model for hadrons in QCD, Phys. Rev. D, Vol. 31, 1985, 2910.

\bibitem{Isgur}
 N. Isgur, R. Kokoski and J. Paton, Gluonic excitations of mesons: Why they are missing and where to find them, Phys. Rev. Lett., Vol. 54, 1985, 869.

\bibitem{Burns}
 T. J. Burns and F. E. Close, Hybrid-meson properties in lattice QCD and flux-tube models,Phys. Rev. D, Vol. 74, 2006, 034003.

\bibitem{Balitsky}
 I. I. Balitsky, D. Diakonov and A. V. Yung, The $0^{++}$ and $0^{-+}$ mass of light-quark hybrid in QCD sum rules, Phys. Lett. B, Vol. 112, 1982, 71.

\bibitem{Chen22}
 H. X. Chen, W. Chen and S. L. Zhu, New hadron configuration The double-gluon hybrid state, Phys. Rev. D, Vol. 105, 2022, 051501.

\bibitem{Chen13}
 W. Chen, R. T. Kleiv, T. G. Steele, B. Bulthuis, D. Harnett, J. Ho, T. Richards and Zhu, Mass spectrum of heavy quarkonium hybrids, J. of High Energy Phys, Vol. 09, 2013, 019.

\bibitem{Chen}
H. X. Chen, N. Su and S. L. Zhu, QCD Axial Anomaly Enhances the $\eta\eta$ Decay of the Hybrid Candidate $\eta_{1}$ Chin. Phys. Lett., Vol 39, 2022, 051201.

\bibitem{Chen14}
 W. Chen, T. G. Steele and S. L. Zhu, Masses of the bottom-charm hybrid bGc states, J. Phys. G Nucl. Part. Phys., Vol. 41, 2014, 025003.

\bibitem{Kumar}
R. Kumar and F. Chand, Series solutions to the N-dimensional radial Schr\"odinger equation for the quark–antiquark interaction potential, Phys. Scr., Vol. 85, 2012, 055008.

\bibitem{abushadey19}
M. Abu-shady and H. M. Fath-Allah, The Effect of Extended Cornell Potential on Heavy and Heavy-Light Meson Masses Using Series Method, J. for Foundations and Appl. of Phys. Vol. 6, 2019, 2.

\bibitem{Barnes05}
T. Barnes, S. Godfrey and E. S. Swanson, Higher Charmonia, Phys. Rev. D, Vol. 72, 2005, 054026.

\bibitem{pdg23}
R. L. Workman, Review of Particle Physics,(Particle Data Group), Prog. Theor. Exp. Phys., Vol.2022, 2022, 083C01.

\bibitem{14110585}
W. Chen, J. Ho, T. G. Steele, R. T. Kleiv, B. Bulthuis, D. Harnett, T. Richards and Shi-Lin Zhu, Proceedings, 30th International Workshop on High Energy Phys. Particle and Astroparticle Phys. Gravitation and Cosmology Predictions, Observations and New Projects (IHEP 2014) Protvino, Russ. 2014, 23.

\bibitem{Nosheen17}
N. Akbar, M. Atif Sultan, B. Masud and F. Akram, Conventional and hybrid Bc mesons in an extended potential model, Phys. Rev. D, Vol. 95, 2017, 074018.


\bibitem{godfrey04}
S. Godfrey, Spectroscopy of Bc Mesons in the Relativized Quark Model Phys. Rev. D, Vol. 70, 2004, 054017.

\bibitem{Davies}
C. T. H. Davies, K. Hornbostel, G. P. Lepage, A. J. Lidsey, J. Shigemitsu and J. H. Sloan, $B_c$ spectroscopy from lattice QCD, Phys. Lett. B, Vol.382, 1996, 131.

\bibitem{Nosheen19}
Nosheen Akbar, Faisal Akram, Bilal Masud and M. Atif Sultan, Conventional and hybrid Bc mesons in an extended potential model, Eur. Phys. J. A, Vol.55, 2019, 82.

\bibitem{Kuti97}
K. J. Juge, J. Kuti and C. J. Morningstar, Gluon excitations of the static quark potential and the hybrid quarkonium spectrum, Nucl. Physics B - Proc. Suppl., Vol. 63, 1998, 326.

\bibitem{2024}
A. Alaakol, S. S. Agaev, K. Azizi and H. Sundu, Mass spectra of heavy hybrid quarkonia and bgc mesons, arive 2403, Vol. 1, 12692.

\bibitem{Nosheen2014}
 N. Akbar, M. Atif, B. Masud and F. Akram, Higher hybrid charmonia in an extended potential model, Phys. Rev. D, Vol. 90, 2014, 054001.

\bibitem{Chen14u}
W. Chen, T. G. Steele and S. L. Zhu, Heavy tetraquark states and quarkonium hybrids, The Universe 2, 2014, 1.

\bibitem{iddir}
F. Iddir and L. Semlala, Heavy Hybrid Mesons Masses, 2006, arXiv hep-ph0611165.

\end{thebibliography}
\end{document}